\documentclass[final]{raa06}           %  For uploading to arXiv and final 
\usepackage{graphicx,times}             %for PS/EPS graphics inclusion, new
\usepackage{natbib}
\usepackage{amssymb,amsmath}
\bibpunct{(}{)}{;}{a}{}{,}
\usepackage[colorlinks=true, citecolor=blue]{hyperref}%
\usepackage{graphicx,kantlipsum,setspace}
\usepackage{caption}
\usepackage{orcidlink} 
\usepackage[T1]{fontenc}
\usepackage{ae,aecompl}

\usepackage{graphics,epsf}
\usepackage{amsmath}                % American Mathematical Society package
\usepackage{amsfonts}               % American Mathematical Society fonts
\usepackage{amssymb}                % American Mathematical Society symbol
\usepackage{epsfig}                 % EPS figures
\usepackage{appendix}
\usepackage{graphicx}
\usepackage{float}
\usepackage{color}
\usepackage{multirow}
\usepackage{colortbl}
\usepackage[para,online,flushleft]{threeparttable}
\usepackage{xcolor}

\hypersetup{citecolor=blue, % color for \cite{...} links
            linkcolor=red, % color for \ref{...} links
            menucolor=blue, % color for Acrobat menu buttons
            urlcolor=blue}  % color for \url{...} links
\newcommand{\cm}{{~\rm cm}}

\newcommand{\km}{{~\rm km}}
\newcommand{\s}{{~\rm s}}

\newcommand{\g}{{~\rm g}}

\newcommand{\erg}{{~\rm erg}}
\newcommand{\yr}{{~\rm yr}}

\begin{document}

\title{Forming a clumpy circumstellar material in energetic pre-supernova activity}

%\footnotetext{$*$ Supported by the National Natural Science Foundation of China.}

%   \subtitle{I. Place Your Subtitle Here}

   \volnopage{Vol.0 (20xx) No.0, 000--000}      %%preserved for Editor. DOn't remove!
   \setcounter{page}{1}          %%starting page, preserved for Editor. DOn't remove!

%\author[0000-0003-0375-8987]{Noam Soker}
%\author[0000-0002-9444-9460]{Dmitry Shishkin}

   \author{Shlomi Hillel, Ron Schreier, Noam Soker\,\orcidlink{0000-0003-0375-8987} 
     % \inst{1}
    }  %\orcid{0000-0003-0375-898}

   \institute{Department of Physics, Technion - Israel Institute of Technology, Haifa, 3200003, Israel;   {\it    shlomi.hillel@gmail.com; ronsr@technion.ac.il,; soker@physics.technion.ac.il}\\
%        \and
%             Full institute address for the third author\\
\vs\no
   {\small Received~~20xx month day; accepted~~20xx~~month day}}

\abstract{
We demonstrate by three-dimensional hydrodynamical simulations of energy deposition into the envelope of a red supergiant (RSG) model the inflation of a Rayleigh-Taylor unstable envelope that forms a compact clumpy circumstellar material (CSM). Our simulations mimic vigorous core activity years to months before a core-collapse supernova (CCSN) explosion that deposits energy to the outer envelope. The fierce core nuclear activity in the pre-CCSN explosion phase might excite waves that propagate to the envelope. The wave energy is dissipated where envelope convection cannot carry the energy. We deposit this energy into a shell in the outer envelope with a power of $L_{\rm wave}= 2.6 \times 10^6 L_\odot$ or $L_{\rm wave}= 5.2 \times 10^5 L_\odot$ for 0.32 year. The energy-deposition shell expands while its pressure is higher than its surroundings, but its density is lower. Therefore, this expansion is Rayleigh-Taylor unstable and develops instability fingers. Most of the inflated envelope does not reach the escape velocity in the year of simulation but forms a compact and clumpy CSM. The high density of the inflated envelope implies that if a companion is present in that zone, it will accrete mass at a very high rate and power a pre-explosion outburst. 
}

\maketitle

% ================================================
\section{Introduction}
\label{sec:Introduction}
% ================================================

The early light curve and spectroscopy of many core-collapse supernovae (CCSNe) indicate the presence of compact circumstellar material (CSM), which refers to CSM with which the explosion ejecta interacts within several days after the CCSN explosion. Two prominent examples are the two relatively close and recent CCSNe SN~2023ixf (e.g., \citealt{Bergeretal2023,  Bostroemetal2023, Bostroemetal2024, Grefenstetteetal2023,  Kilpatricketal2023, Tejaetal2023, VanDyketal2024, Huetal2025, Kumaretal2025}) and  SN~2024ggi (e.g., \citealt{ChenXetal2024, JacobsonGalanetal2024, Pessietal2024, XiangD2024, ZhangJ2024, ChenTWetal2025}). In these two CCSNe, there are no indications of pre-explosion outbursts within tens of years before the explosion (e.g., \citealt{Jencsonetal2023, Soraisametal2023, Neustadtetal2024, Shresthaetal2024}). 
  
An enhanced mass loss rate that starts years to weeks before the explosion, possibly, but not necessarily, accompanied by a pre-explosion outburst, might form a compact CSM (e.g., \citealt{Foleyetal2007, Pastorelloetal2007, Smithetal2010, Marguttietal2014, Ofeketal2014, SvirskiNakar2014, Tartagliaetal2016, Yaronetal2017, Morozovaetal2018, Wangetal2019, Prenticeetal2020, Bruchetal2021, JacobsonGalanetal2022}).
Alternatively, there might exist a long-lived dense CSM around the CCSN progenitor (e.g., \citealt{Dessartetal2017}), as an extended, accelerated zone of the wind (e.g., \citealt{Moriyaetal2017, Moriyaetal2018, MoriyaSingh2024}) or a long-lived circumstellar zone of parcels of gas that are uprising and falling above the stellar photosphere (e.g., \citealt{Soker2021effer, Soker2023Effer, Soker2024ggi, FullerTsuna2024}; see further discussion by \citealt{FullerTsuna2024}). 

The energy source that triggers enhanced pre-explosion stellar activity should be in the core, as only the core rapidly evolves years to days before the explosion. This energy source can be the excitation of waves by the vigorous core convection (e.g., \citealt{QuataertShiode2012, ShiodeQuataert2014, RoMatzner2017, WuFuller2021, WuFuller2022}) or convection-power enhanced magnetic activity in the core (e.g., \citealt{SokerGilkis2017, CohenSoker2024}).  
The energy deposition might lead to substantial envelope expansion that enhances mass loss rate and even leads to some mass ejection and weak to mild outbursts (e.g., \citealt{Fuller2017, OuchiMaeda2019, OuchiMaeda2021}), but cannot by itself trigger an energetic outburst that mimics a weak supernova (e.g., \citealt{McleySoker2014}). A close companion that accretes mass from the inflated envelope and launches jets might power an energetic pre-explosion outburst (e.g., \citealt{Soker2013B, DanieliSoker2019}). 

In some cases, observations indicate that the CSM does not cover the entire sphere, e.g., SN~2023ixf;  (e.g., \citealt{Vasylyevetal2023}). For the CSM of SN~2023ixf, \cite{SmithNetal2023} consider equatorial mass concentration and \cite{Singhetal2024}  and \cite{Bostroemetal2024} mention a clumpy CSM, as the expectation in the effervescent model (e.g., \citealt{Soker2023Effer, Soker2024ggi}) and the similar boil-off model studied by \cite{FullerTsuna2024}.  

In this study, we show that the deposition of large amounts of energy in the envelope results in Rayleigh-Taylor instability (RTI) that leads to the formation of a clumpy CSM. Large amounts refer to possible pre-explosion core activity that deposits energy at a power much larger than the regular stellar luminosity.  
\cite{Fuller2017} performed one-dimensional (1D) simulations and noticed that the energy deposition forms a high-pressure, low-density zone prone to RTI, as \cite{LeungFuller2020} confirmed with 2D simulations. He mentioned that the RTI can smooth the density radial gradient and mix envelope zones.  
The present study explores some properties of the three-dimensional (3D) RTI modes. Section \ref{sec:Methods} describes the 3D hydrodynamical code and its setting and lists our assumptions. Our results are in Section \ref{sec:Results}. In Section \ref{sec:Summary}, we summarize this study.

% ================================================
\section{Methods and assumptions}
\label{sec:Methods}
% ===============================================

% ==============================
\subsection{Numerical code and stellar model}
\label{subsec:Code}
% ==============================

Our numerical setup is as in our previous papers, e.g., \citep{Hilleletal2023, Schreieretal2025}. We built a red supergiant (RSG) stellar model from a zero-age-main-sequence star of metalicity $Z=0.02$ and mass $M_{\rm 1,ZAMS}=15 M_\odot$, evolving it with the 1D stellar evolution code \texttt{MESA} \citep{Paxtonetal2011, Paxtonetal2013, Paxtonetal2015, Paxtonetal2018, Paxtonetal2019}. We transport the 1D RSG stellar model, with a mass of $M_1=  12.5 M_\odot$ and a radius of $R_{\rm RSG}=881\,R_{\odot}$, into the 3D grid of the 3D hydrodynamical numerical code {\sc flash} \citep{Fryxelletal2000}. To save numerical time, we keep an inner sphere with a radius of $R_{\rm inert} = 0.2R_{\rm RSG}= 176\,R_{\odot}$ to be inert; we do not change its structure. We include the gravity of the RSG star at its value when we start the 3D simulations, including the gravity of the inert inner sphere. We use a Cartesian grid with outflow boundary conditions and take the gas to an ideal gas with an adiabatic index of $\gamma = 5/3$, including radiation pressure. 

%We employ an adaptive mesh refinement (AMR) with a refinement criterion of a modified L\"ohner error estimator (with default parameters) on the $z$ component of the velocity.
The center of the RSG is at the origin. The cell size in most of the simulations is $\Delta _{\rm c}=L_{\rm G}/256 = 9.766 \times 10^{11} \cm$, 
for $L_{\rm G} = 250 \times 10^{12} \cm$. 
In one case, for comparison, we have a low resolution of  $\Delta _{\rm c,low}=L_{\rm G}/128 = 1.953 \times 10^{12} \cm$.

% ==============================
\subsection{Energy deposition}
\label{subsec:Energy}
% ==============================

\begin{figure} %[htb!]
\centering
\includegraphics[width=0.44\textwidth]{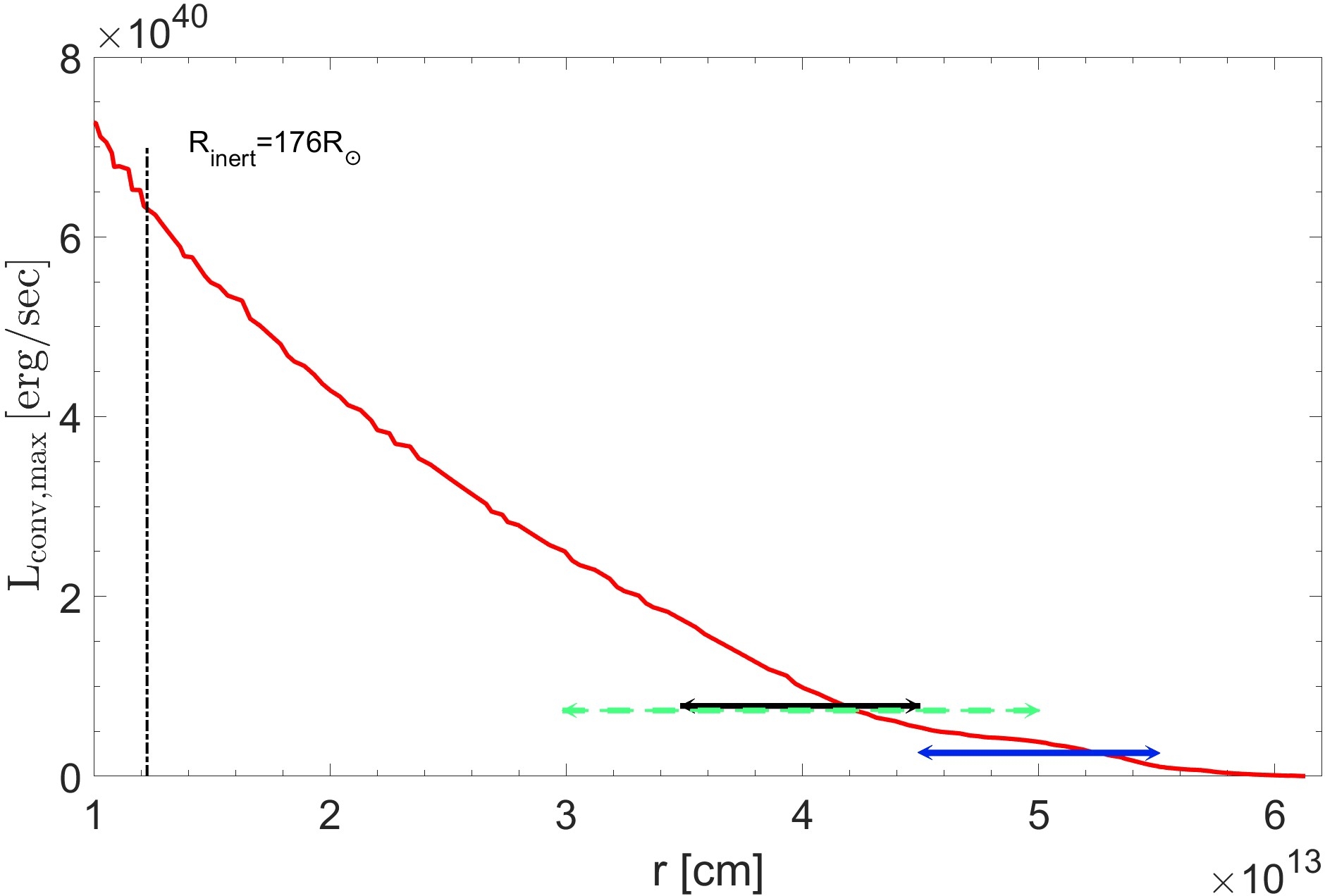}
\caption{ The red line is the maximum power the convection can carry according to equation (\ref{eq:Lconv}) as a function of radius, from $r=10^{13} \cm$ to the edge of the convection zone $R= 6.1 \times 10^{13} \cm$. The vertical line is the outer radius of the inner inert core. The horizontal lines mark our two powers in most simulations, extending over the energy-deposition shell (see Table \ref{Table1}). }
\label{fig:Lcomvmax}
\end{figure}

We consider the pre-explosion process of energy transport from the core to the envelope, e.g., by waves (\citealt{QuataertShiode2012, ShiodeQuataert2014}), as we described in Section \ref{sec:Introduction}. We consider the dissipation of wave energy to occur around the radius where convection at the sound speed is insufficient to carry the wave power $L_{\rm wave}$ (\citealt{QuataertShiode2012, ShiodeQuataert2014}). Namely, where   
\begin{equation}
L_{\rm wave} \simeq L_{\rm conv,max} (r) = 4 \pi r^2 c^3_{\rm S}(r) \rho(r), 
\label{eq:Lconv}
\end{equation}
where $\rho(r)$ is the density and $c_{\rm S}$ the sound speed of the undisturbed stellar model.
In Figure \ref{fig:Lcomvmax} we plot the $L_{\rm conv,max} (r)$ as a function of radius in the 1D stellar model that we use. 

% FFFFFFFFFFFFFFFFFFFFFFFFFFFFFFFFF
% FFFFFFFFFFFFFFFFFFFFFFFFFFFFFFFFF

We simulate mainly for two wave powers, $L_{\rm wave}=1 \times 10^{40} \erg \s^{-1}= 2.6 \times 10^6 L_\odot$ and $L_{\rm wave}=2 \times 10^{39} \erg \s^{-1}= 5.2 \times 10^5 L_\odot$ (for comparison, we simulated two other powers, as we describe in Section \ref{subsec:Others}).
Some 1D studies use similar powers. \cite{McleySoker2014} consider the core-oxygen burning according to the non-rotating stellar model of \cite{ShiodeQuataert2014} and take the wave power, duration, and total energy for the $M_{\rm ZAMS}=15 M_\odot$ stellar model to be $L_{\rm wave, n} = 3.2 \times 10^5 L_\odot$, $t_{\rm wave, n} = 2.3 \yr$, and $E_{\rm wave, n} = 8.9 \times 10^{46} \erg$, respectively. 
The oxygen core burning is shorter than $2.3 \yr$ and somewhat shorter than a year (e.g., \citealt{Fuller2017}). 
\cite{OuchiMaeda2019}, in their 1D simulations, deposited energy with four different powers ranging from $L_{\rm wave}=10^{38} \erg \s^{-1}$ to $L_{\rm wave}=10^{40} \erg \s^{-1}$, starting three years before core collapse. \cite{OuchiMaeda2021} studied the type IIP SN 2009kf, and deposited energy with a power of $L_{\rm wave}=3 \times 10^{39} \erg \s^{-1}$ for three years before core collapse (the explosion). 
 \cite{LeungFuller2020}, in their  benchmark 2D simulation take $L_{\rm wave}=
 3 \times 10^6 L_\odot$ for $t_{\rm wave, n} = 0.33 \yr$.  
 We, therefore, deposit the energy for $t_{\rm wave} \le 10^7 \s= 0.32 \yr$.  

 We inject the energy in two ways. In the uniform injection, the power of the injected energy per unit mass is uniform in the shell and zero outside. Namely, there is a jump at the inner and outer boundaries of the shell in the injected power per unit mass. In the continuously varying injection profile (C-profile), the energy deposition power per unit mass increases linearly from zero at the inner boundary of the shell to its maximum value at the center of the shell (center by radius). Then, it decreases linearly to zero at the outer boundary of the shell.

% ============================
\subsection{Simulations}
\label{subsec:Simulations}
% ============================

We summarize the simulations in Table \ref{Table1}. The first column is the number of the run; the second column gives the inner and outer radii of the shell into which we inject energy, and the third column lists the power of the energy source. The letter C in the second column indicates continuous variation with the radius of the energy deposition power per unit mass (C-profile; Section \ref{subsec:Energy}); otherwise, the energy deposition has a uniform power per unit mass in the energy-injection shell, and sharp decrease to zero outside the shell.
% TTTTTTTTTTTTTTTTTTTTTTTTTTTTTTTTTTTTTTTTTTTTT
\begin{table}
 %\caption{Simulated cases}
%\tiny
%\scriptsize
%\footnotesize
\centering
%\begin{center}
\begin{tabular}{|c|c|c|}
\hline
Run & Shell         & $L_{\rm wave} $   \\
    & $[10^{12} \cm]$ & $ [\erg \s^{-1}]$  \\ 
    \hline
1 & (35-45)  & $10^{40}$  \\ 
2 & (35-45)C& $10^{40}$  \\ 
3 & (30-50)  & $10^{40}$  \\ 
4 & (30-50)C& $10^{40}$  \\ 
5 & (45-55)  & $2\times 10^{39}$  \\ 
6 & (45-55)C& $2\times 10^{39}$  \\ 
7 & (35-45)  & $3.3 \times 10^{40}$  \\ 
8 & (35-45)  & $3.3 \times 10^{39}$  \\ 
\hline
\end{tabular}
 %\end{center}
 %\small 
\caption{The simulated runs, with their number in the first column. In the second column, the two numbers give the inner and outer boundary of the shell into which we inject the energy; C indicates that the energy deposition function is continuous (C-profile; Section \ref{subsec:Energy}), otherwise the energy deposition has a uniform power per unit mass in the energy-injection shell.
The third column lists the energy deposition power.  
}
 \label{Table1}
\end{table}
% TTTTTTTTTTTTTTTTTTTTTTTTTTTTTTTTTTTTTTTTTTTTT

The injection of energy to the shell increases the shell pressure. Consequently, the shell expands and accelerates the layer above it. Because of its expansion, the density in the shell becomes lower than in the layer above it, but its pressure stays larger. This flow is prone to RTIs (e.g., \citealt{LeungFuller2020}). Specifically, if the angle between the density and pressure gradient is larger than $90^\circ$, the typical growth time of the RTI is  
\begin{equation}
  \tau_{\rm RT} \simeq \frac {\rho}{\sqrt{- \overrightarrow{\nabla} P \cdot \overrightarrow{\nabla} \rho}} .
    \label{eq:RTtime}
\end{equation}
To allow the presentation of maps of Rayleigh-Taylor stable and unstable zones, we draw maps of the quantity  
\begin{equation}
f_{\rm st} \equiv \frac {1}{\rho} {\sqrt{\left| \overrightarrow{\nabla} P \cdot \overrightarrow{\nabla} \rho \right|}} ~ \text{sgn} (\overrightarrow{\nabla} P \cdot \overrightarrow{\nabla} \rho ),
    \label{eq:Fst}
\end{equation}
which is a frequency in stable zones (hence the letter `f' and the subscript `st'). 
In unstable zones $f_{\rm st}<0$, and $-1/f_{\rm st}$ is approximately the growth time of the RTI, assuming the wavelength of the perturbation is about the density scaleheight, $d \ln \rho / dr$. In stable zones $f_{\rm st}$ is approximately the Brunt–V\"ais\"al\"a frequency, assuming the density gradient is much steeper than the pressure gradient, $d \ln \rho /dr \gg d \ln P/dr$; the latter holds in our case where the low-density zone are much hotter.

% ================================================
\section{Numerical results}
\label{sec:Results}
% ===============================================
% ============================
\subsection{Run 1}
\label{subsec:Run1}
% ============================

\begin{figure*} %[htb!]
\centering
\includegraphics[width=0.42\textwidth]{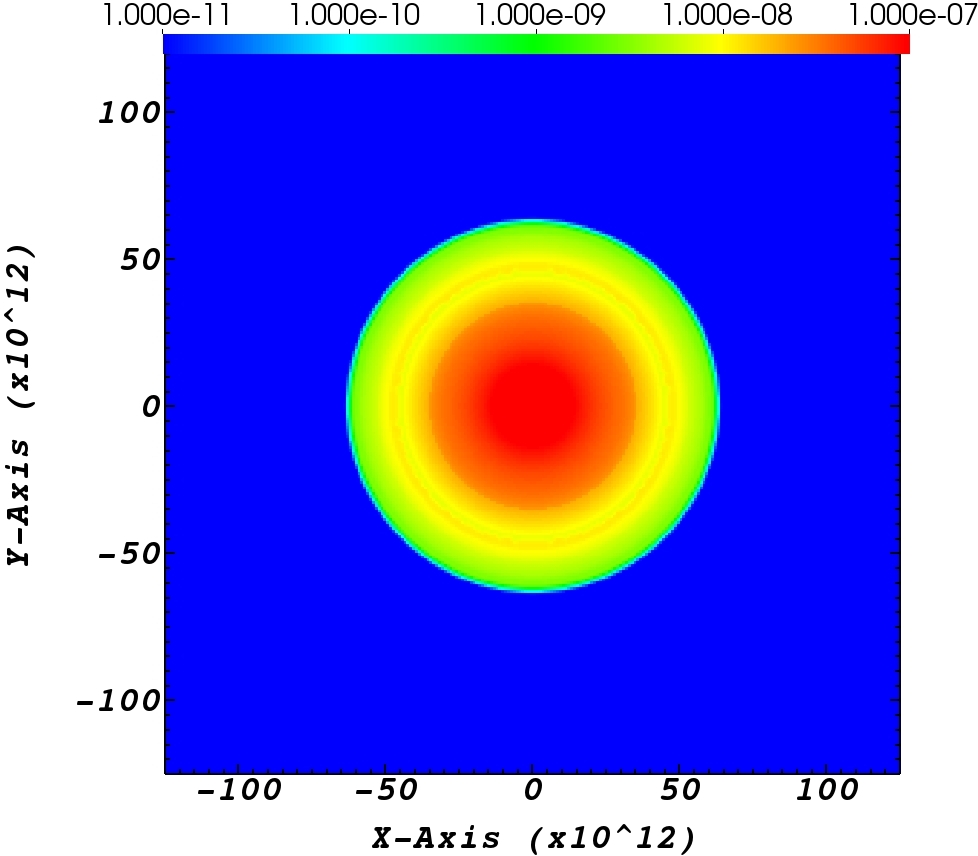}
\includegraphics[width=0.42\textwidth]{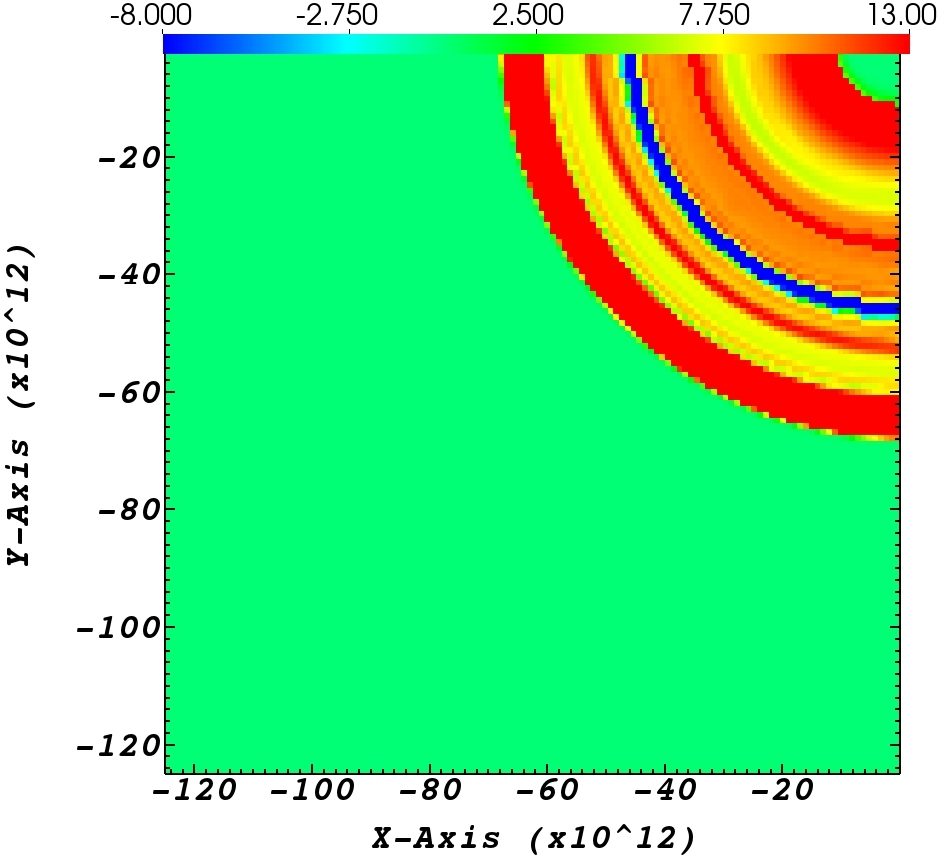}\\
\includegraphics[width=0.42\textwidth]{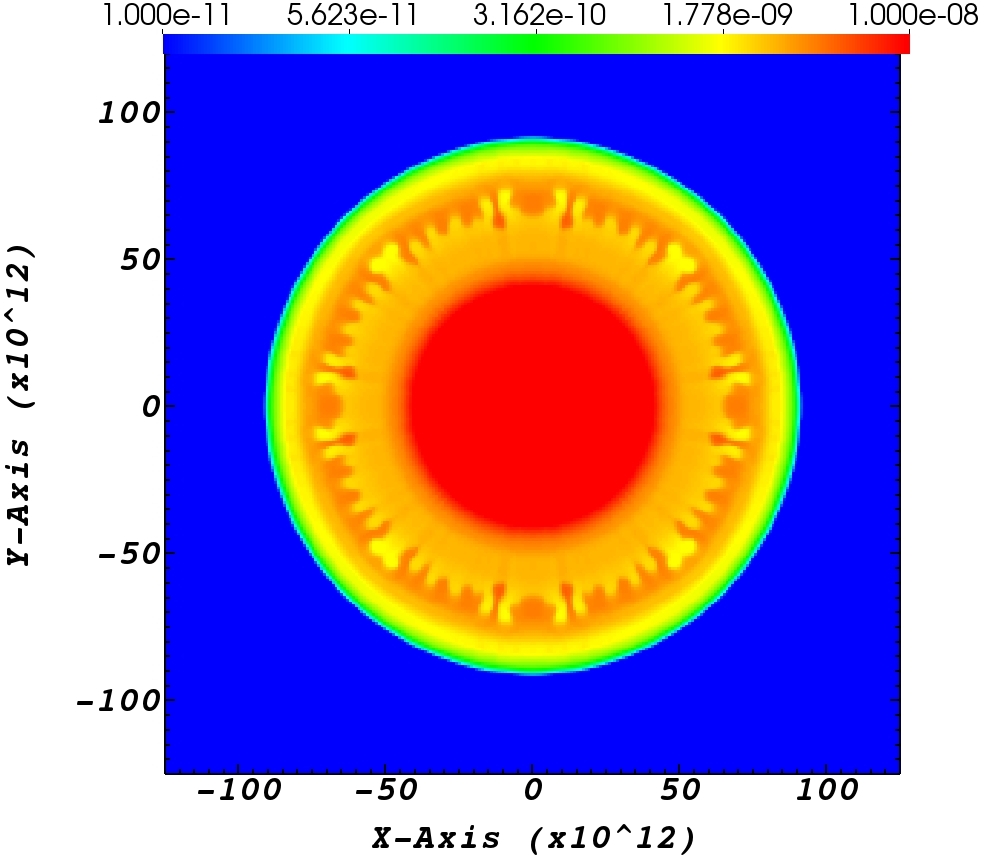}
\includegraphics[width=0.42\textwidth]{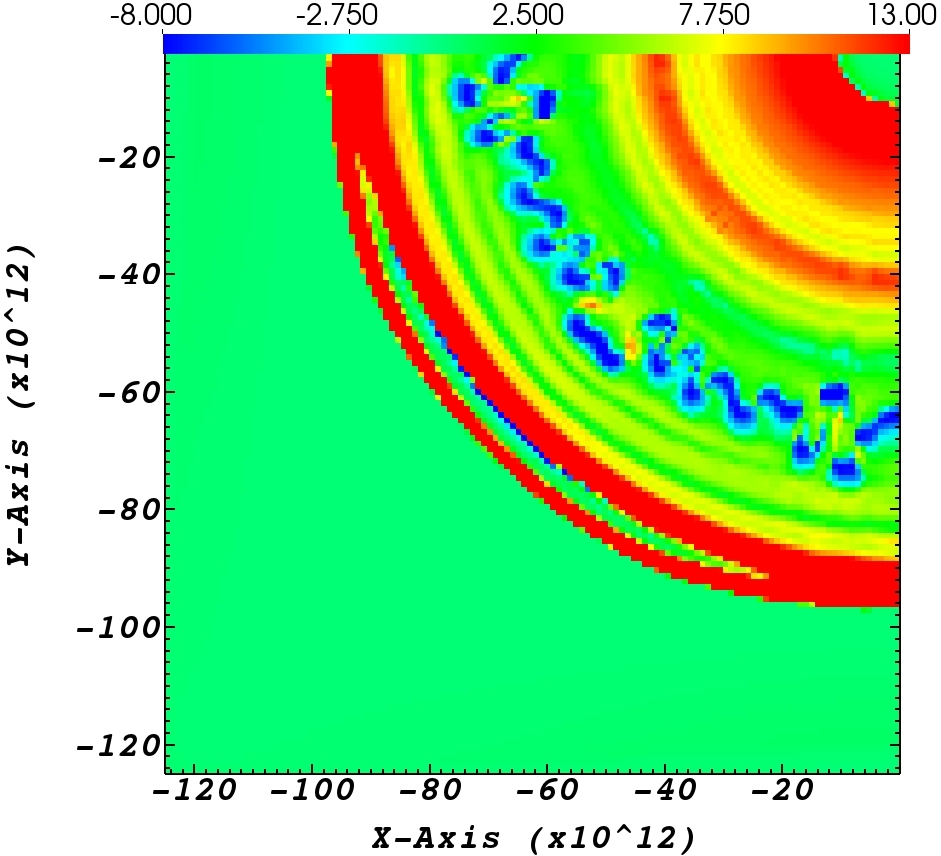}\\
\includegraphics[width=0.42\textwidth]{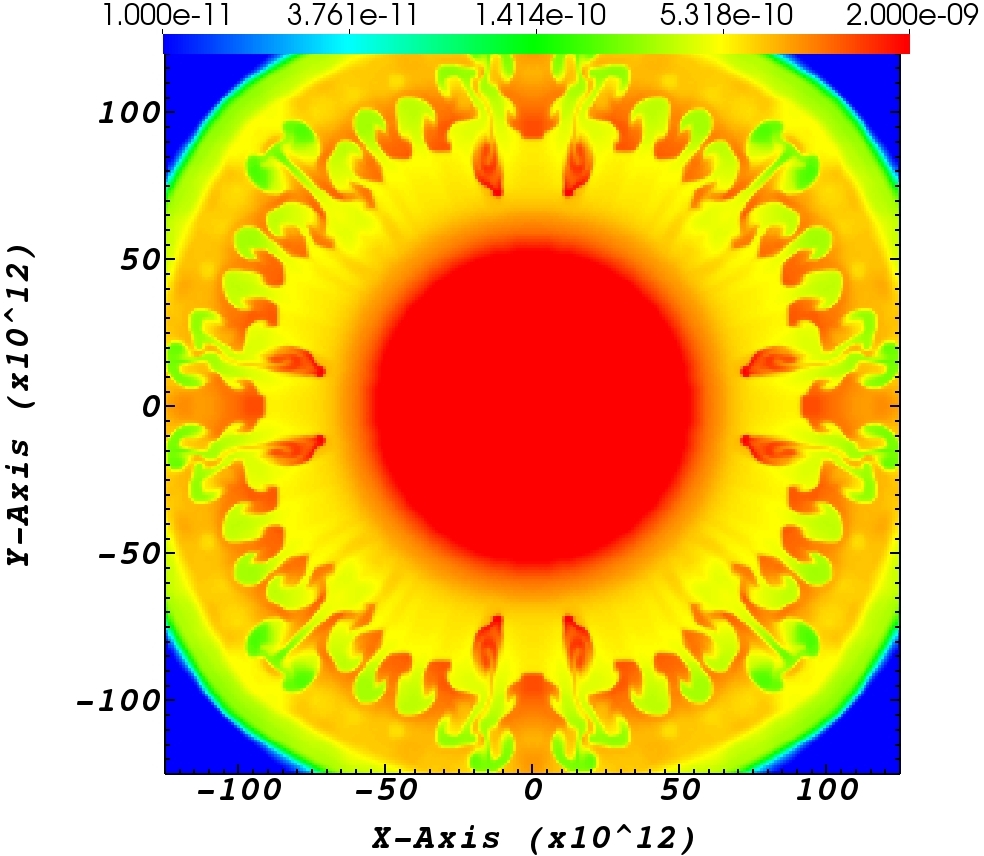}
\includegraphics[width=0.42\textwidth]{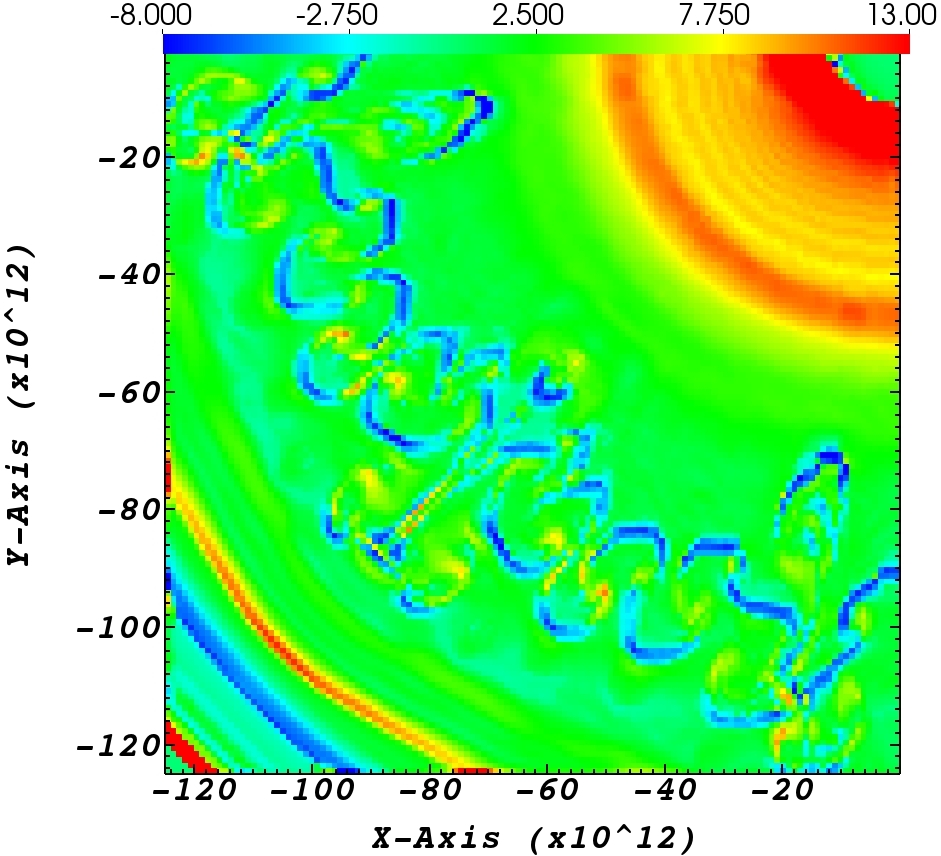}
\caption{Maps of the density (left column, units according to the color bar in $\g \cm^{-3}$) and $f_{\rm ST}$ (equation \ref{eq:Fst}; right column, units in $\yr ^{-1}$ according to the color bars) in the plane $z=0$ and at three times of Run 1 (see Table \ref{Table1}). The times, from top to bottom, are, 
$t_1= 3 \times 10^6 \s = 0.095 \yr$, $t_2= 1.5 \times 10^7 \s = 0.475 \yr$, and $t_3=3 \times 10^7 \s = 0.95 \yr$. 
The RTI maps show only a quarter of the $z=0$ plane. 
Zones with positive values are stable, and $f_{\rm st}$ is about the Brunt–V\"ais\"al\"a frequency. Zones with negative values are unstable, and $-f_{\rm st}$ is the growth rate, i.e., $-f^{-1}_{\rm st}$ is the growth time of the instability. 
}
\label{fig:Run1DenRTI}
\end{figure*}

In this section, we describe Run 1. 
Figure \ref{fig:Run1DenRTI} presents density maps (left column) and maps of $f_{\rm st}$ (right column) in the plane $z=0$ at three times of Run 1.  
The deep blue zones in the RTI maps have a growth time of about $1/8 \yr$, much shorter than the simulation time of almost a year. The two late density maps show the development of the RTI models to the non-linear regime. The RTI forms clumps in the outer envelope and ejecta. Qualitatively, our results are similar to the benchmark simulation of \cite{LeungFuller2020} who have $L_{\rm wave}= 3 \times 10^6 L_\odot$ for $t_{\rm wave, n} = 0.33 \yr$, similar to what we have in Run 1, but they use a different energy deposition scheme and their 2D simulations resolution is lower than ours 3D simulations. Like  \cite{LeungFuller2020}, we find the RTI to be in the non-linear regime after a year, and the RTI fingers and mushrooms to be inside the expanding photosphere. Namely, there is a smooth inflated envelope layer above the highly-non-linear instability clumps.    
% FFFFFFFFFFFFFFFFFFFFFFFFFFFFFFFFF
% FFFFFFFFFFFFFFFFFFFFFFFFFFFFFFFFF

\begin{figure} %[htb!]
\centering
\includegraphics[width=0.42\textwidth]{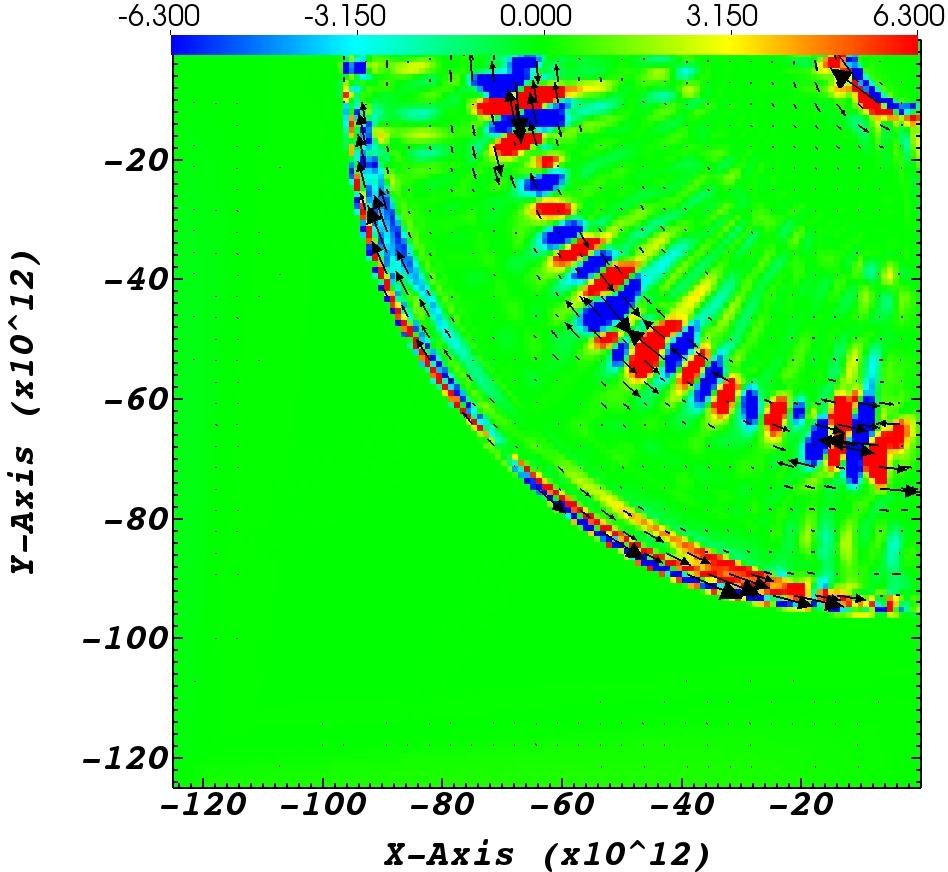}\\
\includegraphics[width=0.42\textwidth]{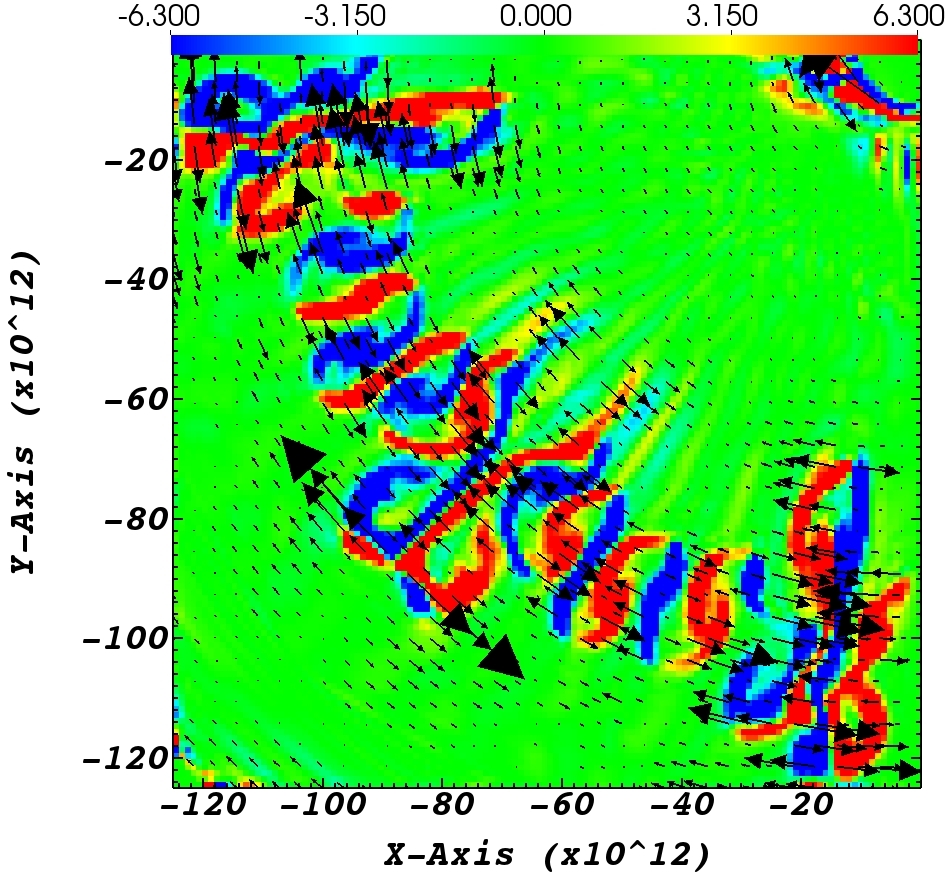}
\caption{The vorticity $(\overrightarrow{\nabla} \times \overrightarrow{v})_z$, in units of $\yr^{-1}$ according to the color bar, in the $z=0$ plane of Run 1 at the two late times, $t_2=  0.475 \yr$, and $t_3= 0.95 \yr$  defined in Figure \ref{fig:Run1DenRTI}. The arrows show the tangential component of the velocity in the $z=0$ plane, with the longest arrows showing a velocity of $v_\theta=117 \km \s^{-1}$.      
}
\label{fig:Vortices1}
\end{figure}

The flow due to the penetration of RTI fingers from dense to lower density and vice versa, forms vortices. Figure \ref{fig:Vortices1}  presents vorticity of the flow in the $z=0$ plane, i.e., the quantity $(\overrightarrow{\nabla} \times \overrightarrow{v})_z$ (in units of $\yr^{-1}$). 
The areas in the plane $z=0$ that show large RTI mushrooms (lower-left panel of Figure \ref{fig:Run1DenRTI}) coincide with the area of high vorticity (lower panel of \ref{fig:Vortices1}). The typical circularization timescale, $(\vert \overrightarrow{\nabla} \times \overrightarrow{v} \vert)^{-1}$, of the strongest vortices is much shorter than the simulation time. The vortices, like the RTI mushrooms, are well-developed after about a year.   
% FFFFFFFFFFFFFFFFFFFFFFFFFFFFFFFFF
% FFFFFFFFFFFFFFFFFFFFFFFFFFFFFFFFF

\begin{figure} %[htb!]
\centering
\includegraphics[width=0.42\textwidth]{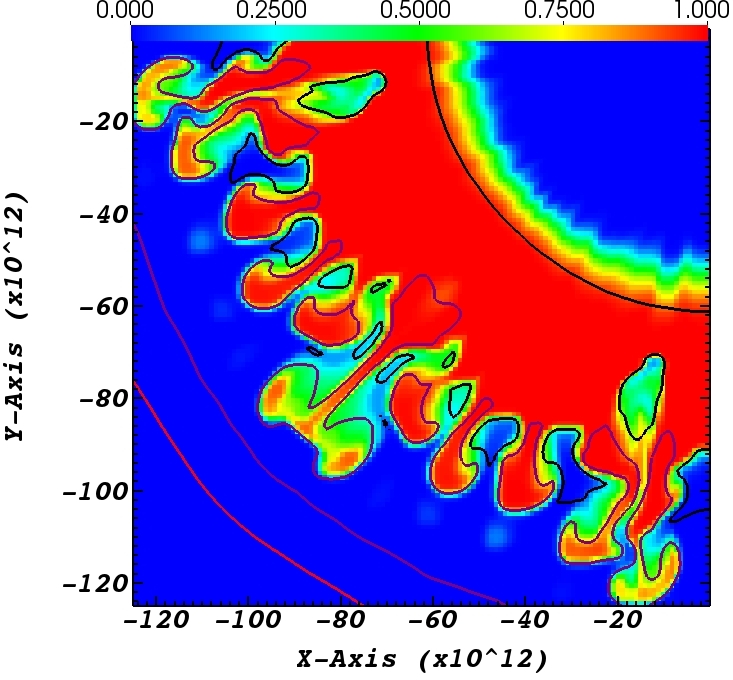}\\
\includegraphics[width=0.42\textwidth]{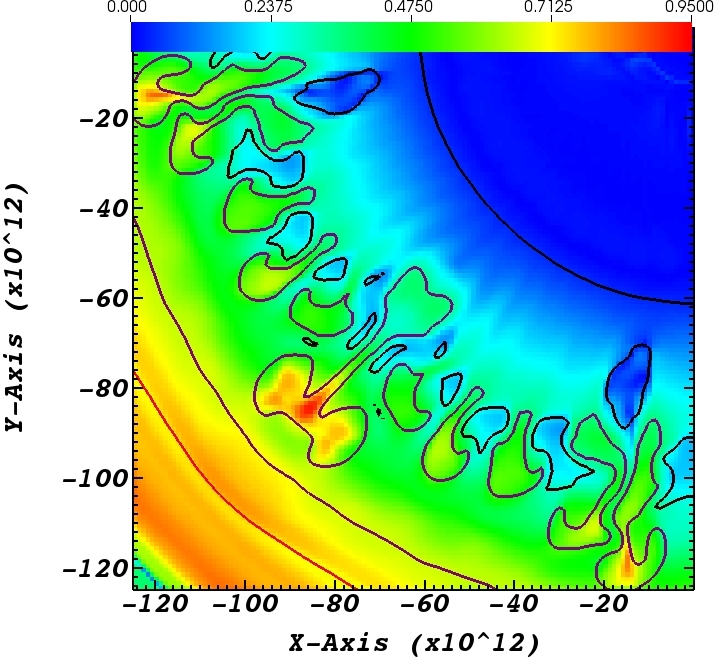}
\caption{Two quantities of Run 1 in one quarter of the $z=0$ plane at $t_3=0.95 \yr$. Upper panel: The tracer of the shell into which we inject the energy, from zero (blue) to 1 (red). The tracer follows the mass distribution that started in the shell $35 \times 10^{12} - 45 \times 10^{12} \cm$. The tracer map shows that the energy-deposition shell material (in green to red colors) occupies the low-density zones and emphasizes the RTIs. Lower panel: The ratio of the magnitude of the velocity to the escape velocity, according to the color bar, from zero (deep blue) to 0.95 (deep red). The three density contours are red: $\rho=3 \times 10^{-11} \g \cm^{-3}$ (the most outer contour only); brown: $\rho=5 \times 10^{-10} \g \cm^{-3}$; black:  $\rho=10^{-9} \g \cm^{-3}$ (the innermost contour and the small islands).      
}
\label{fig:TracerSpeed}
\end{figure}

We further emphasize the RTI mushrooms in Figure \ref{fig:TracerSpeed}. 
In the upper panel, we present the tracer map of the energy-deposition shell. The tracer is a numerical quantity that follows the flow of a designated volume. In this case, the tracer of the gas in the shell $35 \times 10^{12} < r < 45 \times 10^{12} \cm$ into which we deposit energy is given an initial value of 1. With time, the value of the tracer in a cell shows the fraction of mass that started in the shell; the tracer can have any value from 0 to 1. A comparison of the upper panel of Figure \ref{fig:TracerSpeed} with the lower-left quarter of the lower-left panel of Figure \ref{fig:Run1DenRTI} shows that the original material in the shell occupies the low density volume. In the lower panel of Figure \ref{fig:TracerSpeed}, the colors depict the ratio of the magnitude of the velocity to the escape velocity at that location. There are also three density contours on that panel.  At the edge of the grid $r=125 \times 10^{12}\cm=1800 R_\odot$ the escape velocity is $52 \km \s^{-1}$. The ejecta have not reached the escape velocity. Our result is similar to \cite{LeungFuller2020} who also find that in their benchmark simulation at most a small mass in the outer layer reaches the escape velocity. In any case, the explosion is expected to occur within the year we simulate, or shortly thereafter, before much mass is ejected \citep{LeungFuller2020}. 
% FFFFFFFFFFFFFFFFFFFFFFFFFFFFFFFFF
% FFFFFFFFFFFFFFFFFFFFFFFFFFFFFFFFF

\begin{figure} %[htb!]
\centering
\includegraphics[width=0.42\textwidth]{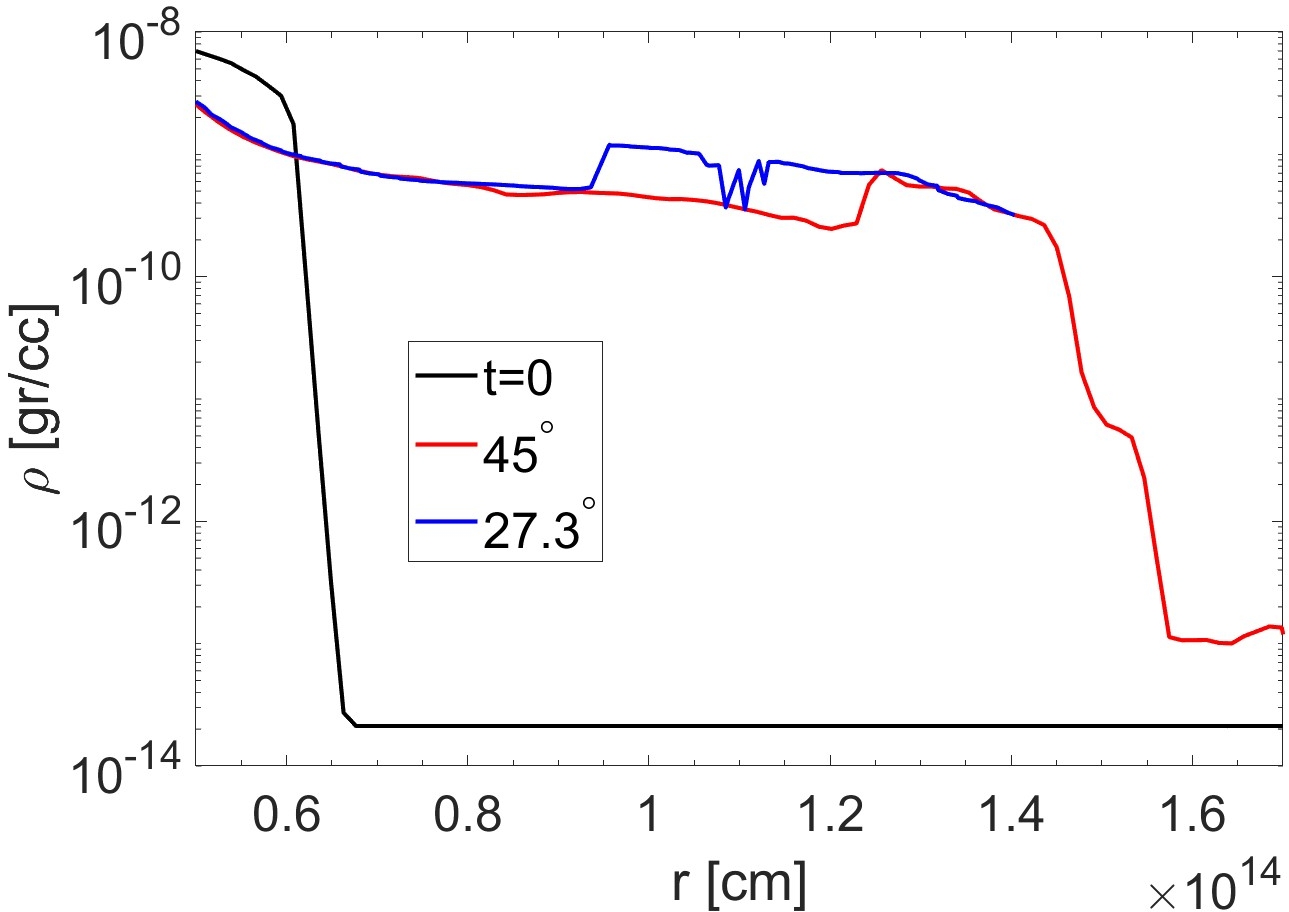}
\caption{Density profiles of simulation Run 1. The black line shows the density profile of the initial model. The red line shows the density profile of Run 1 at  
$t_3=0.95 \yr$ along the diagonal line $x=y$ in the first quarter of the plane $z=0$. The blue line is at the same time and plane, but a line from the center at $27.3^\circ$ to the $x$-axis, a line that goes through dense RTI mushrooms.        
}
\label{fig:DensityProfile1}
\end{figure}

Within about a year from the beginning of the energy deposition, which is the burning of the core oxygen, we expect the star to explode. Before that, a possible close companion might accrete mass from the expanding ejecta and launch jets that power a much brighter event months to weeks before the explosion. Indeed, the ejecta density is high to a distance of $\simeq 2000 R_\odot$,  more than twice the initial RSG radius. We present the density profile of the outer envelope and the ejecta in Figure \ref{fig:DensityProfile1}. In addition to the initial profile (black line), we show the density profile along a diagonal which has no dense RTI clumps (red line), and along another direction that contains RTI clumps and is denser (blue line). 
If a companion within the ejecta exists, it can accrete at a very high rate, launch jets, and power a pre-explosion outburst. Because the ejecta is optically thick, this binary interaction is like a common envelope evolution. For a density of $\rho = 10^{-10} \g \cm^{-3}$, the Bondi-Hoyle-Lyttleton accretion rate for a companion of $1.4M_\odot$ orbiting at a radius of $1600R_\odot$ is $\dot M_{\rm acc} \approx 0.01 M_\odot \yr^{-1}$. A main sequence star of this mass releases an accretion gravitational energy at a power of $\approx 10^5 L_\odot$, while a neutron star releases energy at a power of $\approx 10^{10} L_\odot$. Due to the negative jet feedback mechanism (e.g., \citealt{GrichenerCohenSoker2021, Hilleletal2022Feedback}), these powers will be lower by more than an order of magnitude; in particular, the power of the neutron star jets will be lower by two to three orders of magnitude from that above power \citep{Hilleletal2022Feedback}. The powering of a pre-explosion outburst by an accreting companion is the subject of a future study.      
% FFFFFFFFFFFFFFFFFFFFFFFFFFFFFFFFF
% FFFFFFFFFFFFFFFFFFFFFFFFFFFFFFFFF

% ============================
\subsection{Other cases}
\label{subsec:Others}
% ============================

We describe the results of our other simulations as density maps in the plane $z=0$ at $t_3=0.95 \yr$, except Run 7.   
Run 1 is the one we studied in detail in Section \ref{subsec:Run1}. Run 1L has the same physical parameters as Run 1, but the numerical resolution is lower because the grid cell size is twice as large as in the other simulations. As expected, the RTI fingers and mushrooms are larger but reach the same level of non-linearity (similar to the results of the low-resolution 2D simulations of \cite{LeungFuller2020}). The seven other simulations have different physical initial parameters. We describe them by order.  
% FFFFFFFFFFFFFFFFFFFFFFFFFFFFFFFFF
\begin{figure*} %[htb!]
\centering
\includegraphics[width=0.32\textwidth]{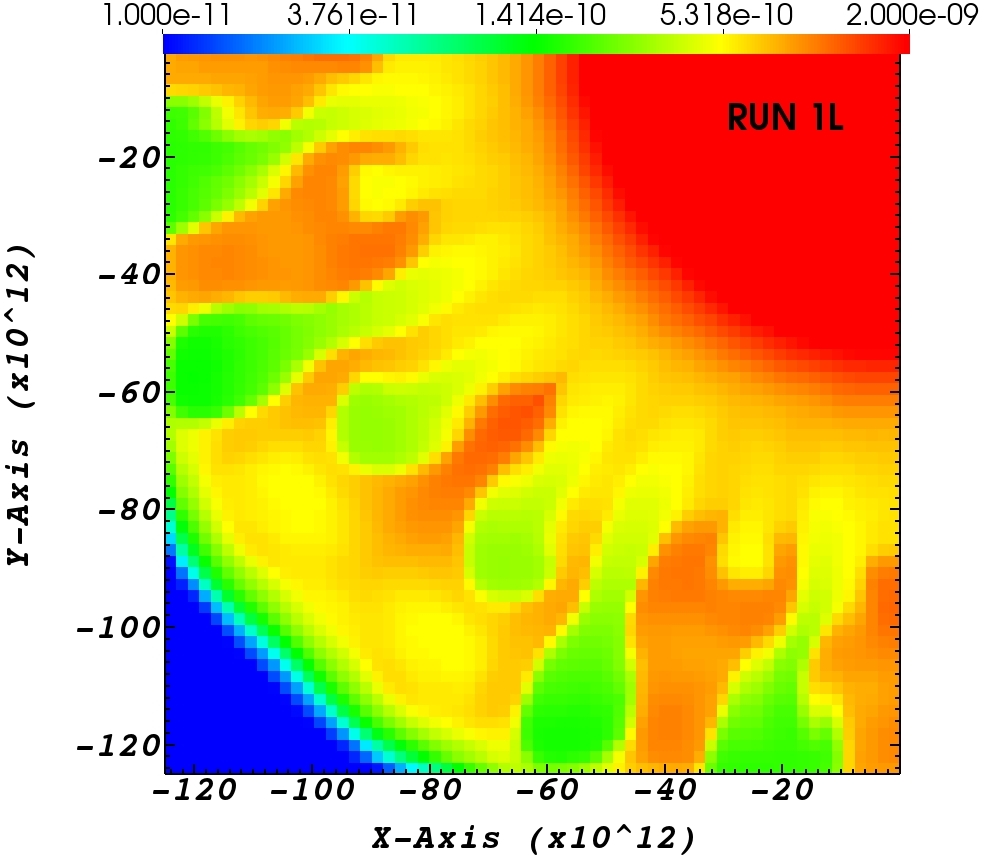}
\includegraphics[width=0.32\textwidth]{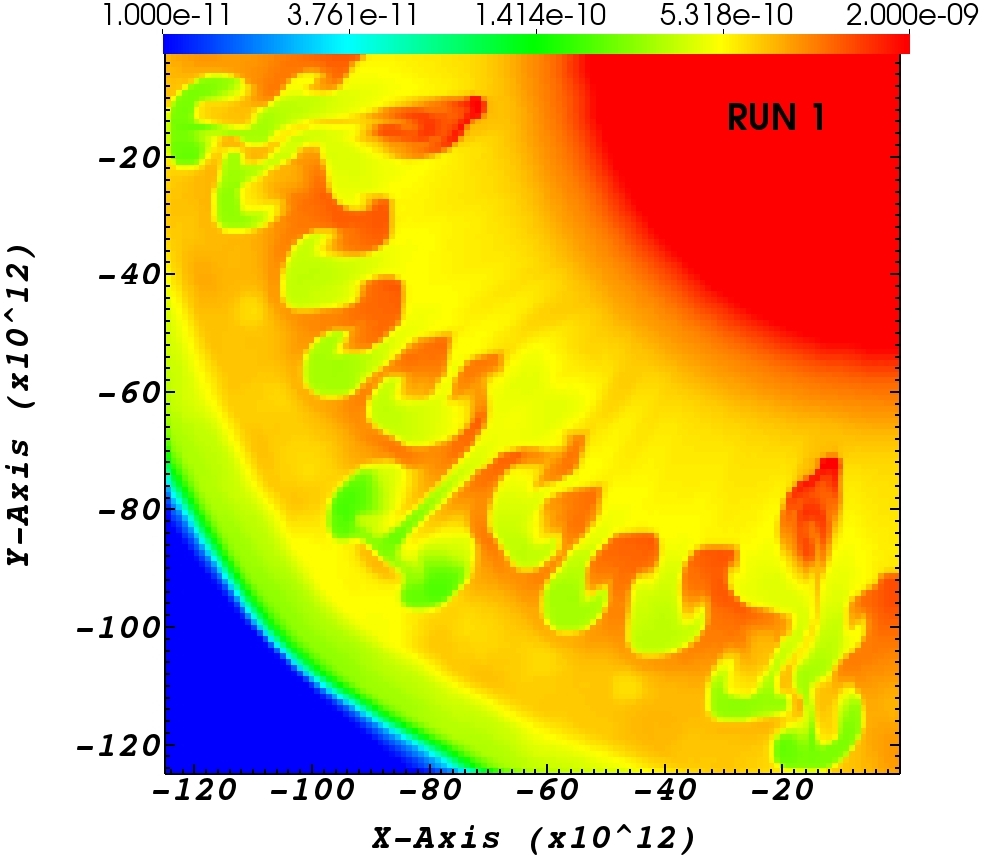}
\includegraphics[width=0.32\textwidth]{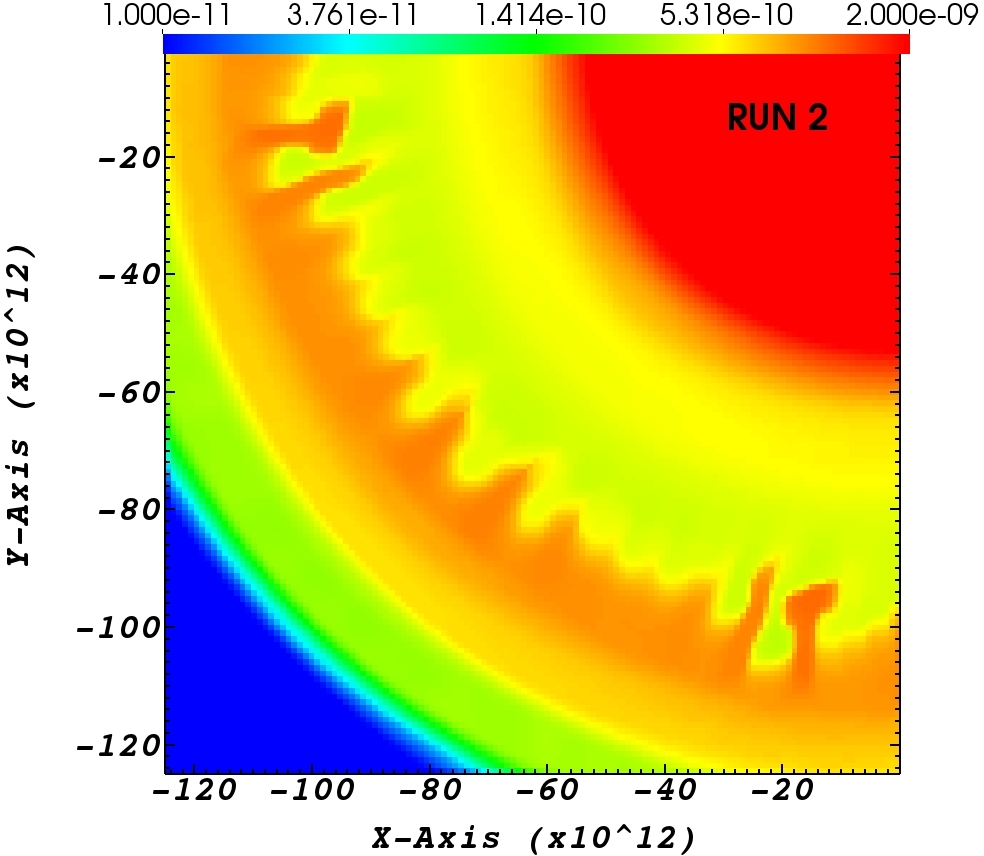}\\
\includegraphics[width=0.32\textwidth]{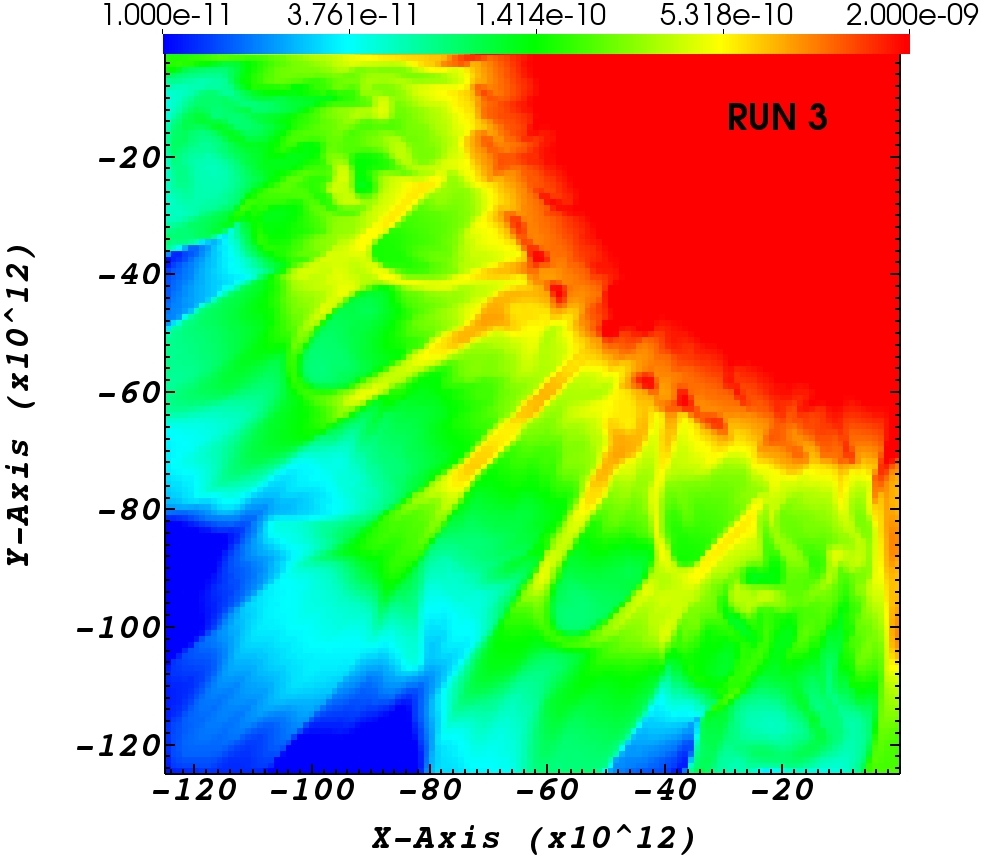}
\includegraphics[width=0.32\textwidth]{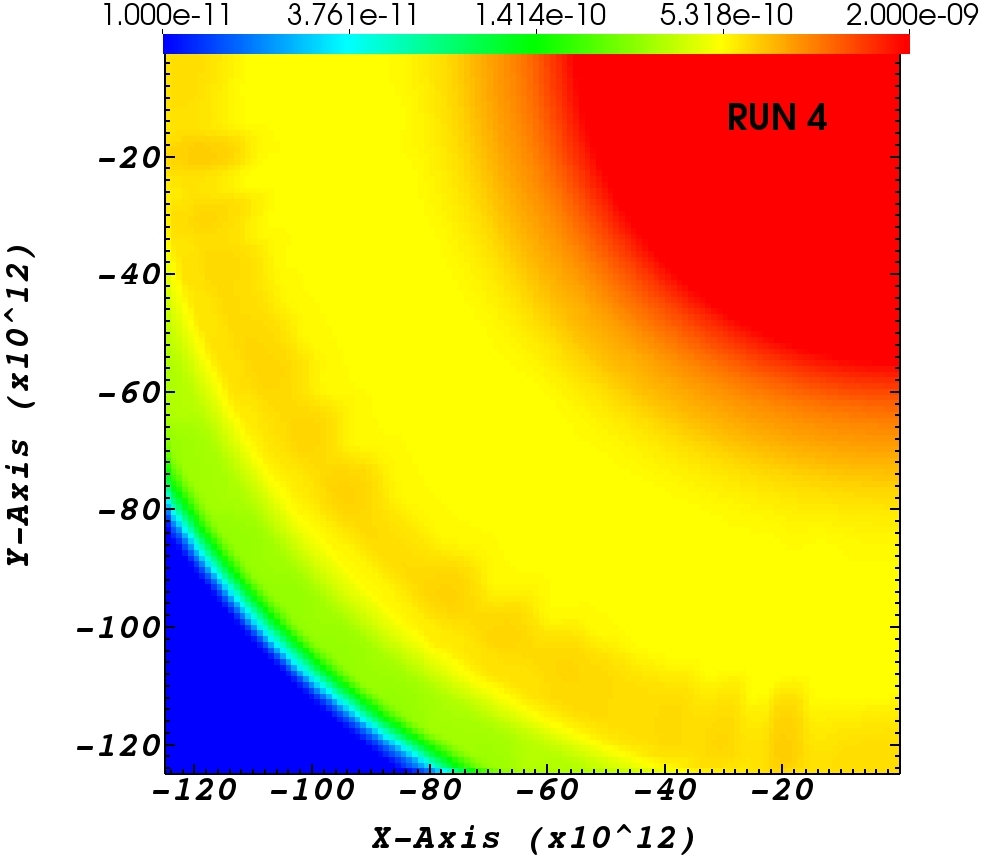}
\includegraphics[width=0.32\textwidth]{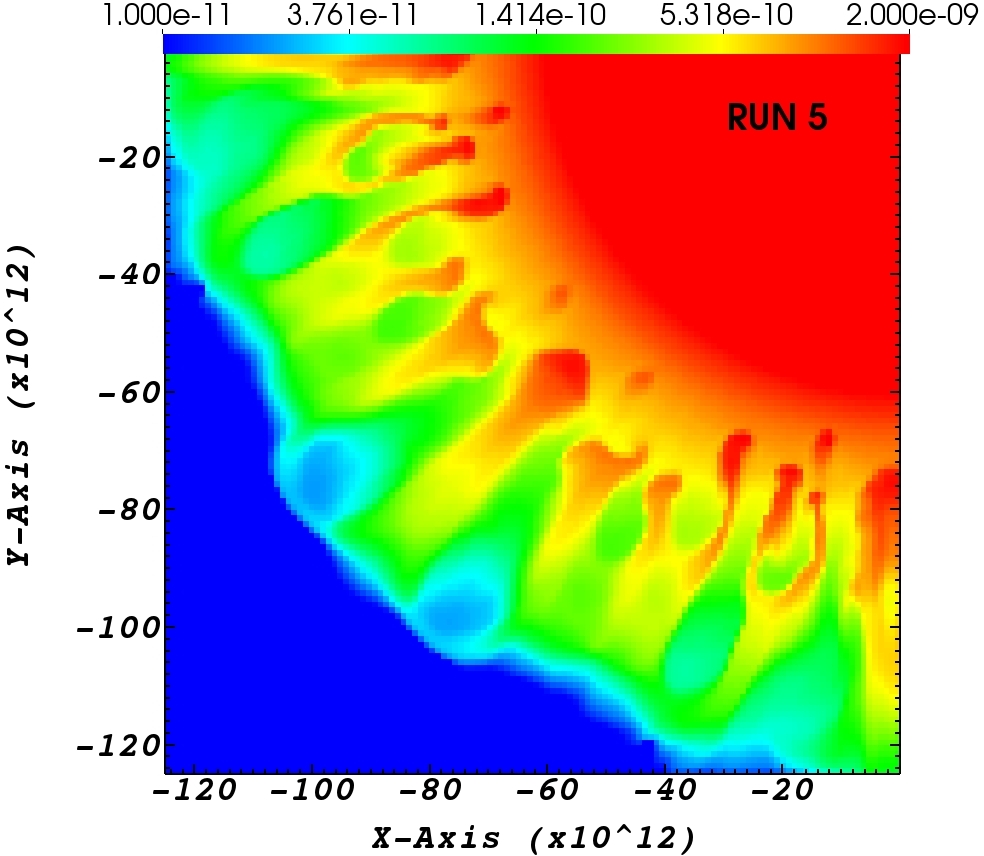}\\
\includegraphics[width=0.32\textwidth]{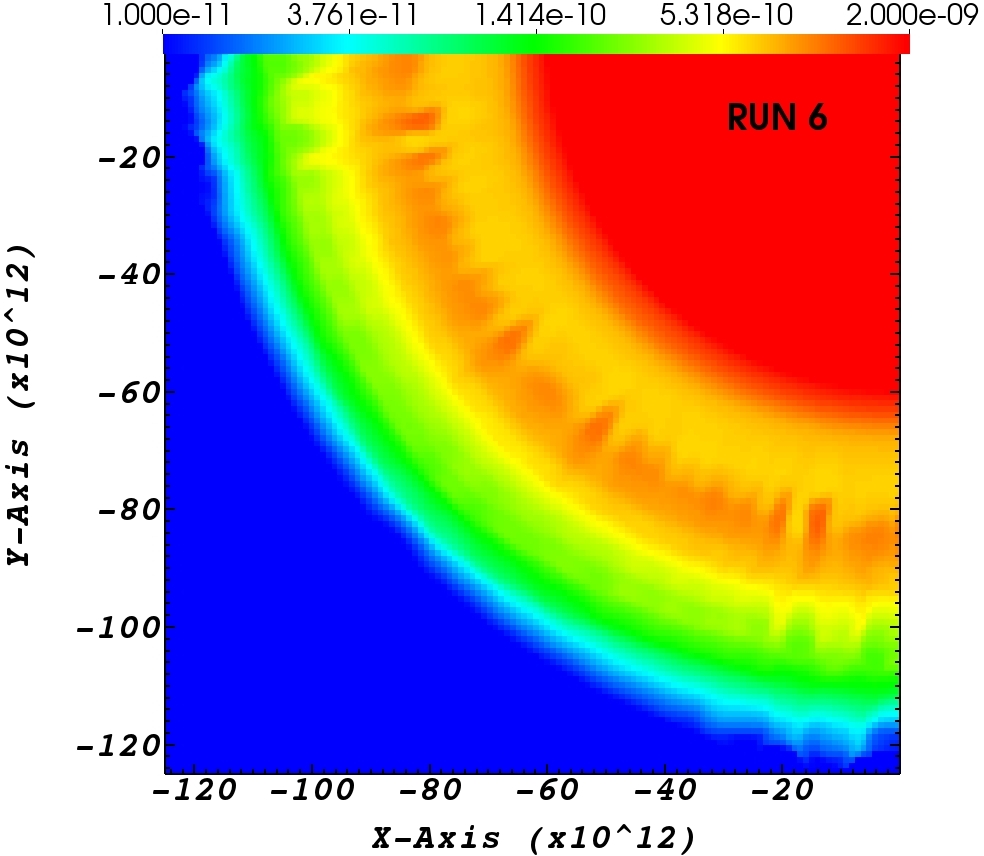}
\includegraphics[width=0.32\textwidth]{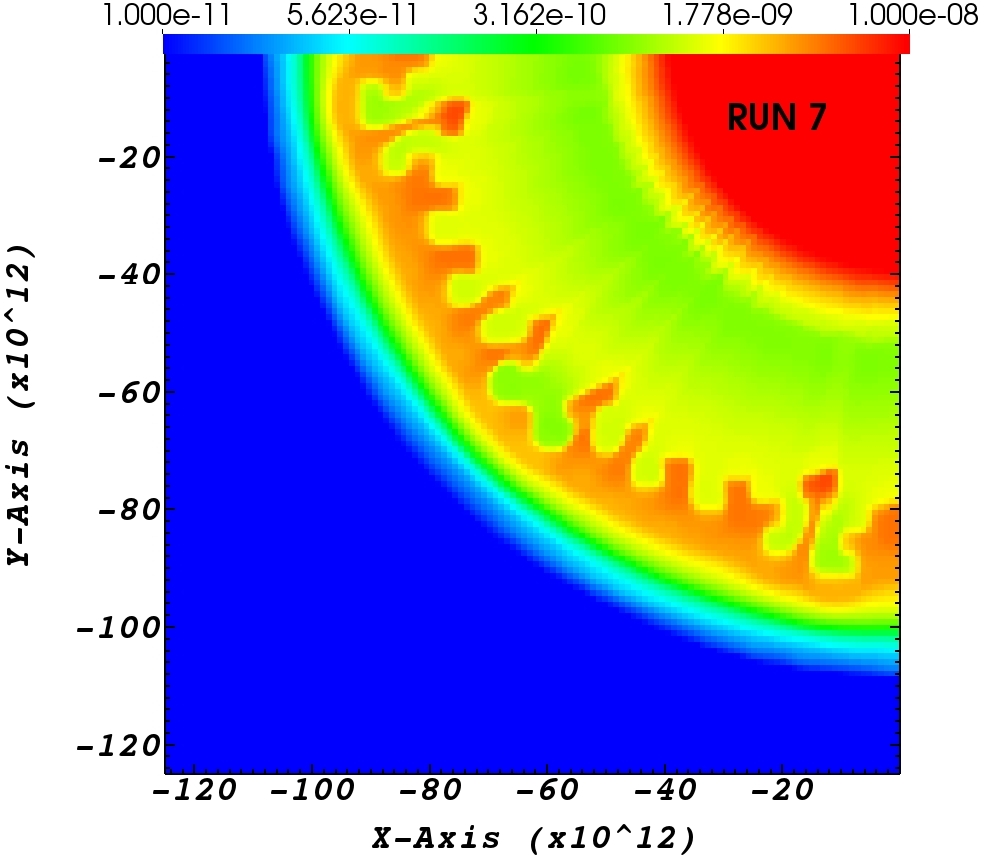}
\includegraphics[width=0.32\textwidth]{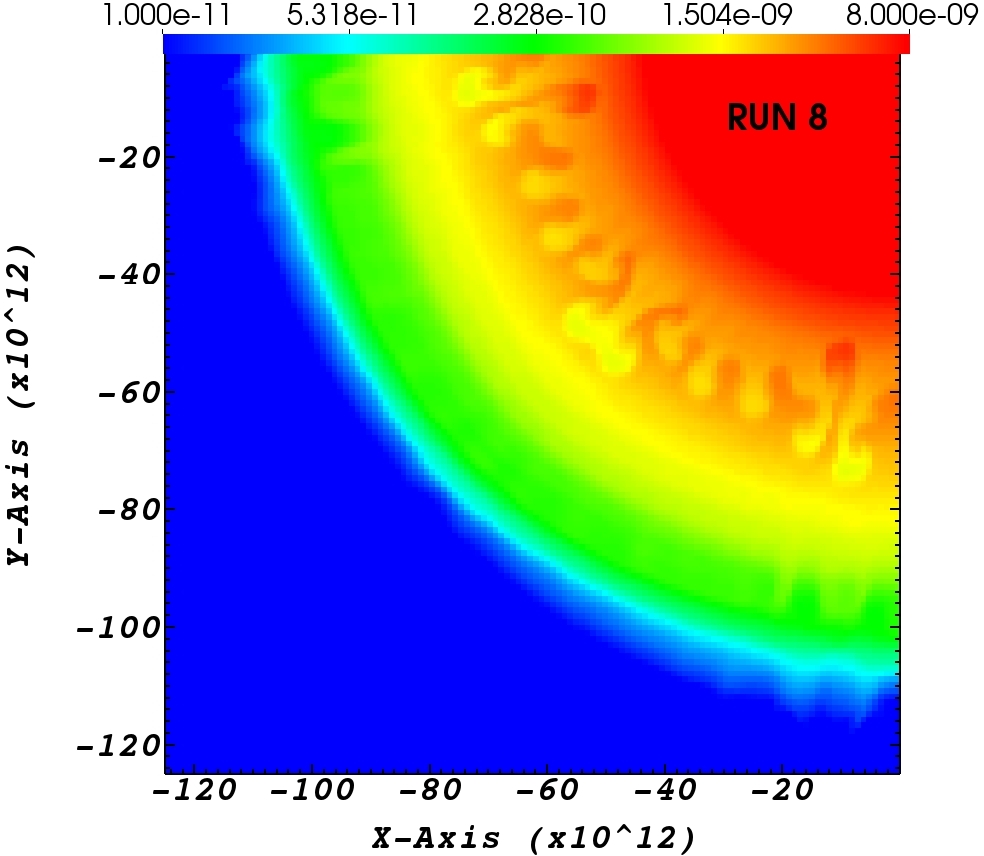}
\caption{Density maps in the plane $z=0$ of nine simulations, RUN 1L (low resolution), and the regular resolution RUN 1 to RUN 8
(see Table \ref{Table1}). All maps are at $t_3=3 \times 10^7 \s = 0.95 \yr$, except RUN 7 which is at at $t=1.25 \times 10^7 \s = 0.4 \yr$. All the simulations developed RTI fingers and mushrooms by one year. 
}
\label{fig:Density_all_runs}
\end{figure*}
% FFFFFFFFFFFFFFFFFFFFFFFFFFFFFFFFF

Run 2 differs from Run 1 in that the energy power per unit mass increases linearly from zero at the inner shell's boundary to a maximum at the center of the shell. It decreases linearly to zero at the shell's outer edge (C-profile); there is no sharp jump in the energy deposition power at the shell boundaries. The total jet's power is the same as in Run 1. The same type of C-profile is also in Run 4 and Run 6. The shock front (the jump from blue to green color in the lower left corner of the panels in Figure \ref{fig:Density_all_runs}) reaches the same radius in Run 1 and Run 2. The largest difference is that the RTI developed slower in Run 2. There are clear RTI fingers at $t_3=0.95 \yr$, but not yet well-developed RTI mushrooms. Because of their lower resolution, it is hard to compare the simulations of \cite{LeungFuller2020} to ours. From inspecting their low-resolution RTI fingers development, it seems that their energy deposition scheme yields RTI development in between our Run 1 and Run 2. We therefore suggest that the real situation will be in between Run 1 and Run 2, namely, between the sharp-jump energy deposition profile and the C-profile.   
  
Run 3 differs from Run 1 in the energy deposition shell which is twice as wide in radius. The outer edge reaches much closer to the photosphere in mass coordinate, and there is not much mass above the energy deposition shell. The RTI figures break into the low-density outer envelope and become columns rather than mushrooms. Run 4 is like Run 3, but the energy deposition has a C-profile (no sharp jumps in the energy deposition per unit mass at the boundaries of the shell). We can observe the RTI figures only at the beginning of their non-linear phase; their density is about twice that of their surroundings. 

Runs 5 and 6 are the low-energy outburst with energy deposition in the outer envelope (see Table \ref{Table1}). Although the power is $20 \%$ of that in Run 1 - Run 4, the mass involved is much smaller, and the RTI fingers reach the non-linear regime. In Run 5, with a sharp boundary, the RTI develops much larger fingers, similar to that in Run 3. In Run 6, with the C-profile of energy deposition, the fingers are at an earlier phase of evolution at $t_3=0.95 \yr$.   

We close the comparison simulations with Run 7, which is similar to Run 1 but is three times as powerful as Run 1 and Run 8, which is one-third of the power of Run 1 (Table \ref{Table1}). The RTI of Run 7 develops much faster than in Run 1 and the envelope expands much faster (as \citealt{LeungFuller2020} found), such that at $t_3=0.95 \yr$ the RTI fingers and mushrooms are outside the numerical grid. Therefore, we present the density map at $t=0.4 \yr$. Clear RTI fingers and mushrooms appeared at that time. As expected, Run 8 expands slower than Run 1 because of its lower energy. Despite that, the RTI fingers and mushrooms are well developed at $t_3=0.95 \yr$.      

All the simulations show the development of RTI fingers and mushrooms in the expanding envelope. This forms a clumpy and inhomogeneous CSM.   

% ================================================
\section{Summary}
\label{sec:Summary}
% ===============================================

We conducted 3D hydrodynamical simulations of large energy deposition into the envelope of an RSG star to mimic the energy deposition that the fierce nuclear burning in the core might release years to months before the explosion (Section \ref{sec:Introduction}). We concentrated on cases with two energy deposition powers (Figure \ref{fig:Lcomvmax}), and added two extra powers for comparison (Table \ref{Table1}; Figure \ref{fig:Density_all_runs}). We changed the energy deposition shell and the profile of energy deposition. 
The energy deposition inflates the envelope (Figures \ref{fig:Run1DenRTI} and \ref{fig:Density_all_runs}). Most of the mass does not reach the escape velocity in our simulations (e.g., the lower panel of Figure \ref{fig:TracerSpeed}), but a fraction might escape. Some mass does not escape but reaches close to the escape velocity; hence, it will rise to large distances. We did not follow the ejecta beyond twice the initial stellar radius of $R_{\rm RSG}=881 R_\odot$. Our results are similar to the 2D simulations of \cite{LeungFuller2020}, but with our 3D simulations, we much better resolve the development of the RTI. We also cover some other parameters than they do, although our Run 1 has similar physical parameters to their benchmark case.  

Our main result is that the prominent inflated envelope and ejected mass that result from large energy deposition to the outer envelope form a clumpy CSM, as all figures show, due to RTI (right column of Figure \ref{fig:Run1DenRTI}). 

Observations of some CCSNe that interact with a compact CSM also suggest that the CSM might be clumpy, e.g., SN 2023ixf (e.g., \citealt{Singhetal2024, Bostroemetal2024}). 
However, this does not necessarily imply that the compact CSM of SN 2023ixf or other CCSNe with clumpy CSM is due to energy deposition. There are no indications for a pre-explosion outburst in SN 2023ixf (e.g., \citealt{Jencsonetal2023, Neustadtetal2024, Soraisametal2023} ).
An alternative to the energy deposition is that the large amplitude pulsation and vigorous convection in the RSG star progenitor lift bound material to several, and even tens of, stellar radii, as in the effervescent model (e.g., \citealt{Soker2023Effer, Soker2024ggi}) and the similar boil-off model \citep{FullerTsuna2024}. \cite{Freytagetal2024} find in their simulation that the pulsation and convection of AGB stars lead to episodic levitation of dense gas clumps due to the RTI; No extra energy deposition exists in their AGB model. 
 
We conclude that the compact CSM of CCSNe is expected to be highly clumpy, whether formed by stellar pulsation and convection or by the deposition of core energy to the outer envelope. 

Our simulation shows that the energy deposition does not accelerate much mass to escape velocities. It forms a compact, dense, extended envelope (Figure \ref{fig:DensityProfile1}).  
A companion accreting from this extended envelope/CSM, whether a main sequence star, a neutron star, or a black hole, can release gravitational energy (Section \ref{subsec:Run1}), mainly via jets, that can eject mass and power a bright pre-explosion outburst (theory, e.g., \citealt{Soker2013B, DanieliSoker2019} and observation, e.g., \citealt{Pastorelloetal2025}).
This interaction will introduce a large-scale departure from spherical symmetry. One goal of CCSN observations shortly after the explosion should be to search for departures from sphericity due to clumps (short scale) and binary interaction (large scale). Note that because of the clumpy medium that the companion accretes, the angular momentum of the accreted material might vary, and the jets the companion launches might have varying directions, wobbling jets (e.g., \citealt{Dorietal2023}), as suggested for multi-polar planetary nebulae \cite{AvitanSoker2025}.  

% ===================================================
\begin{acknowledgement}
% ===================================================
A grant from the Pazy Foundation supported this research. 
\end{acknowledgement}

\end{document}